\documentclass{optica-article}

\journal{opticajournal} % for journals or Optica Open

\articletype{Research Article}

\usepackage{lineno}
\usepackage{gensymb}
\usepackage{fancyhdr}
\usepackage[nolist]{acronym}
\usepackage{xspace}

\begin{acronym}
    \acro{SAS}{saturated absorption spectroscopy}
    \acro{CQED}{cavity quantum electrodynamics}
    \acro{FPC}{Fabry-P\'erot cavity}
    \acrodefplural{FPC}{Fabry-P\'erot cavities} % needs special handling for long plural form
    \acro{FOV}{field of view}
    \acro{PD}{photodiode}
    \acro{PC}{plano-concave}
\end{acronym}

\newcommand{\distA}[1]{%
  Approved for public release; distribution is unlimited.  Public Affairs %
  release approval %
  AFRL-2026-0088.
}

\newcommand{\mm}{\thinspace\ensuremath{\mathrm{mm}}\xspace}     % millimeters
\newcommand{\um}{\thinspace\ensuremath{\mathrm{\mu m}}\xspace}  % microns
\newcommand{\nm}{\thinspace\ensuremath{\mathrm{nm}}\xspace}     % nanometers
\newcommand{\GHz}{\thinspace\ensuremath{\mathrm{GHz}}\xspace}   % GigaHertz
\begin{document}

\title{Enhanced stability from co-resonant cavities in a monolithic array}

\author{Alexandra Crawford,\authormark{1,*} Jacob Williamson,\authormark{1} Robert Leonard,\authormark{3} Ce Pei,\authormark{1} Aniruddha Bhattacharya,\authormark{1} Meagan Plummer,\authormark{2} Seth Hyra, \authormark{4} Spencer Olson,\authormark{4} and Chandra Raman\authormark{1}}

\address{\authormark{1}School of Physics, Georgia Institute of Technology, 837 State Street, Atlanta, Georgia 30332, USA\\
\authormark{2}Universities Space Research Association, Washington, DC 20024, USA\\
\authormark{3}Space Dynamics Laboratory, Quantum Sensing \& Timing, North Logan, UT 84341, USA\\
\authormark{4}Air Force Research Laboratory, Kirtland Air Force Base, NM 87117, USA\\

}

\email{\authormark{*}alexandra.p.crawford1@gmail.com} %% email address is required; see note below about the corresponding author designation

% use {asbstract*} to suppress the copyright line. Copyright information will be added in production

%==================================================================

\begin{abstract*} 

We demonstrate a micro-Fabry-P\'erot cavity array through laser etching of high-surface-quality mirrors onto a single fused silica substrate.  A cavity finesse of $4750\pm 200$ was achieved with a simple array design with $500 \um$ cavity length, $100 \um$ diameter micromirrors and $300 \um$ transverse separation.  Arrays with up to 12 cavities were simultaneously tested for single mode operation, and absolute frequency measurements correlated strongly with the etched depth as measured by profilometry.  Simultaneous measurements of the absolute resonant frequency for neighboring cavities showed a factor of 5 common-mode cavity drift reduction.  Arrays of such cavities can be employed in chip-scale cavity QED networks (current cooperativity estimates are at the border of strong coupling for $^{87}$Rb atoms, $C=1$) as well as for precise laser stabilization at nearby wavelengths on a chip.

\end{abstract*}

%%%%%%%%%%%%%%%%%%%%%%%%%%  body  %%%%%%%%%%%%%%%%%%%%%%%%%%
\section{Introduction}

There is a growing interest in  microfabrication of optical resonators for applications such as \ac{CQED} and precision metrology.  For cavity QED, the historical approach has focused on the interaction between one cavity and one atom as a testbed system \cite{kimble1998_strong}.  The scalability afforded by planar microfabrication allows us to imagine entirely new possibilities where arrays of nearly identical resonators function in concert to produce non-classical states of light and novel regimes of multi-particle cavity QED.  To achieve this end, it is essential to develop platforms that allow easy access to neutral atoms or ions.  

\begin{figure}[htbp]
\centering\includegraphics[width=\textwidth]{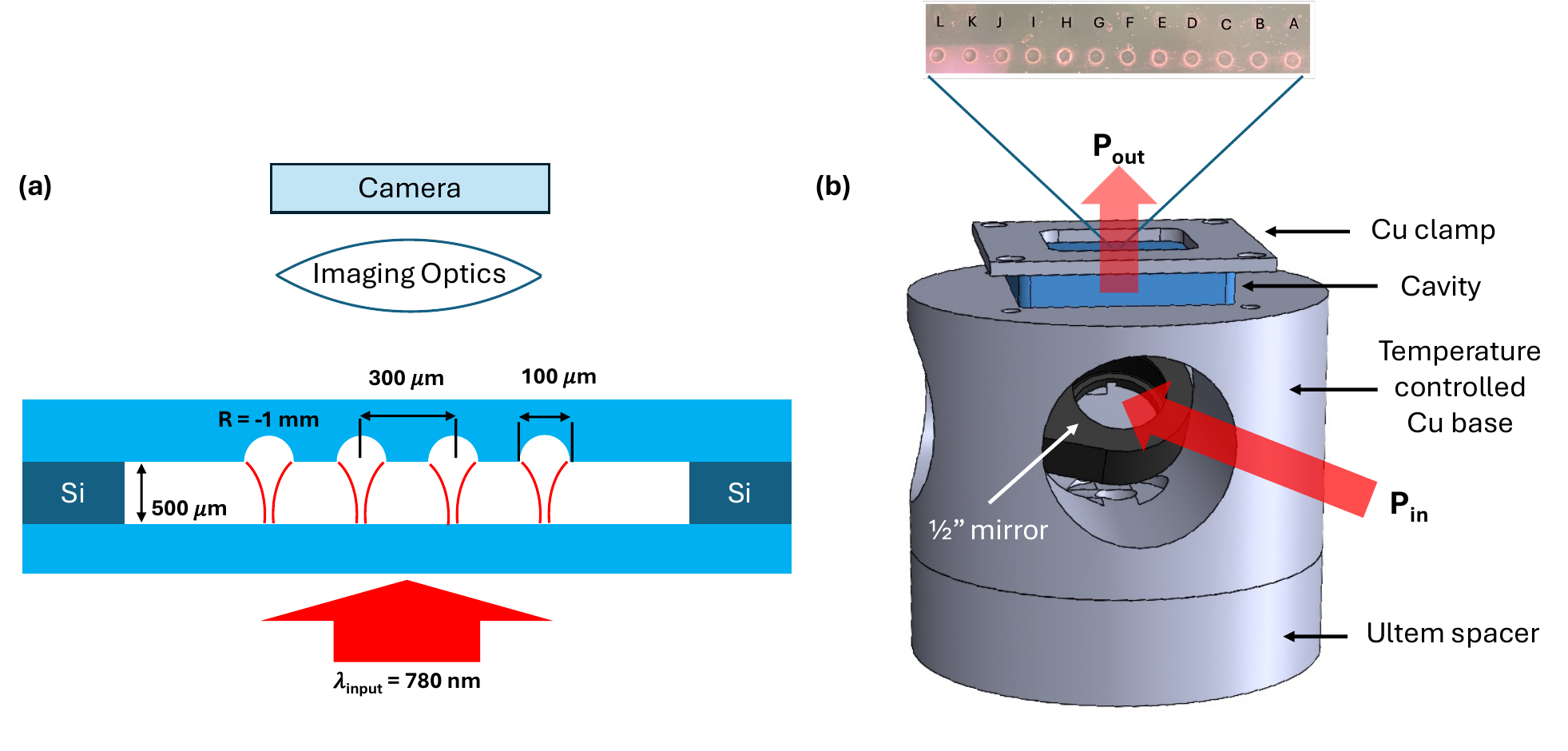}
\caption{(a) Assembled \ac{PC} cavity array uses silicon as a spacer material between the two fused silica substrates. Each etch in the array has a $-1 \mm$ radius of curvature ($\mathcal{R}$), $100 \mm$ diameter, and are spaced $300 \um$ apart center to center. Imaging optics and a detector collects cavity transmission. (b) Cavity array mounted flat on a heated copper (Cu) base is clamped down with a square Cu clamp. Ultem spacer is used to thermally isolate the mount from optical table. Input light is routed vertically through the cavity array. The naming convention of the cavities in the array follow a letter assignment as seen in the zoomed-in picture of the mounted cavity array.}
\label{fig:setupMount}
\end{figure}

Electron-beam lithography (EBL) is a mature method that has enabled highly scalable integrated planar nanophotonic systems \cite{eshaghian_2020highQ}. The evanescent field from light coupled into an EBL-fabricated structure extends into free-space, enabling atom-light coupling.  While the planar design promises manufacturing scalability, a significant disadvantage with EBL-based on-chip resonators is the limited extent of the mode field, extending no further than about $0.5 ~\mu$m from the chip surface.  This makes coupling to three-dimensional atom-based systems, such as laser traps or thermal atom beams, very challenging.  In contrast, a structure which allows for mode field propagation {\em normal} to the photonics plane allows strong atom-light coupling up to millimeter distances from a chip surface. This structure may be fabricated using spatially separated planar structures, thereby maintaining the ease of manufacture of a planar design, such as the micro-\ac{FPC} structure shown in Fig.~\ref{fig:setupMount}a.

\par

In this paper, we have experimentally demonstrated the concept of a micro-FPC array, and fully characterized the absolute frequencies of its various resonator elements. We utilized CO$_2$ laser ablation, a technique that can create high quality concave depressions \cite{hunger_fiber_2010, najer_fabrication_2017, uphoff_frequency_2015,gallego_high-finesse_2016, doherty_multi-resonant_2023}, to build an array of 12 cavities on a fused silica substrate with a single spacer element. Our approach enables scalable cavity architectures with the potential to easily integrate with, for example, a chip-scale collimated atom beam source \cite{li_2019cascaded ,larsen_chip-scale_2025}. A key challenge in implementing arrays of resonators is that resonance frequency may vary dramatically across the array.  These differences in resonance frequency arise due to small uncontrolled variations during the fabrication process. In this work we perform an analysis of passive relative stability in the fabricated cavity array, which were not deliberately created to have the same resonant frequency. To the best of our knowledge there have been no measurements of the relative frequency stability of adjacent microcavities fabricated on a monolithic substrate nor demonstrations of the passive suppression of environmental noise in such platforms.  Achieving passive stability of the absolute resonator frequency across the array is a critical requirement for scaling up many quantum applications.  Prior work required active mounting mechanisms to bring adjacent cavities into co-resonance \cite{doherty_multi-resonant_2023}.  These added degrees of freedom to cavity mounting increase mechanical complexity, which makes scaling difficult.

%==================================================================

\section{Fabrication and Setup}

Micro-cavity arrays were fabricated by milling into polished fused silica substrates (ESCO Optics Q110125) using a $\mathrm{CO_2}$ iterative refinement laser etching process. Building on our work from Plummer et al. \cite{plummer_2025iterative}, we applied this milling technique to achieve the desired dimensions and low surface roughness for each micro-cavity. This feedback-controlled process, which repeatedly measures and corrects the surface shape through successive material removal, was optimized to pattern large arrays of concave depressions. The depressions have a transverse center-to-center separation of $300 \um$ (Fig. \ref{fig:setupMount}a), each with a diameter of $100 \um$ and a nominal $-1 \mm$ radius of curvature. Consistent with our previous work, we performed phase shifting interferometry (PSI) profilometry measurements and analysis on the micro-milled features. The analysis yielded two key findings: an achieved RMS surface deviation of 2--$3 \nm$ over the entire surface of each sphere and a measured radius of curvature of approximately $\mathcal{R} =-1 \mm$.

As discussed in \cite{plummer_2025iterative}, the cavity finesse is dominated by the central Gaussian beam region on the scale of the beam spot, for which the inferred finesse is typically an order of magnitude higher than that obtained when considering the full mirror surface. The surface roughness provides an upper bound to the achievable cavity finesse. Our measured cavity finesse values are well within the scattering-limited estimate, and are likely limited by the reflectivity of the mirrors.

\par
Prior to assembling the cavities, a high reflectivity coating ($99.9\%$) was applied to the entire mirror array from an outside vendor.  After applying coatings, a hemispherical cavity array was formed by placing a milled cavity array opposite a flat mirror. The cavity was mounted using a clamping plate, while silicon spacers set the nominal cavity length. Silicon provides a balance between good passive stability and ease of manufacture, while spacers with lower CTE such as ULE glass \cite{saavedra_2021tunable} would improve the stability even further. The silicon spacers have a nominal thickness of $500\um$. The length of the cavity was varied via material thermal expansion. Foam padding was added to the bolts pushing the clamp against the base to improve passive mechanical stability. A drawing of the cavity mount is shown in Fig.~\ref{fig:setupMount}b. A copper base was heated with a $25.4 \mm \times 50.8 \mm$ Omega polyimide heater \mbox{(PLMLV-101/10)} and Omega \mbox{CN32PT-220} temperature controller. The temperature of the mount was monitored to a precision of $\pm 0.1 \degree$C. An Ultem spacer was used to thermally isolate the base from the optical table. The platform was designed to allow for stable cavity mounting while allowing transmission measurement of the cavities. Light was routed through the cavity array by a half-inch mirror mounted inside the heated base. Cavity resonance modes were measured at a fixed temperature by scanning the input laser frequency while measuring transmission.

%==================================================================

\section{Experiment and Results}

%==================================================================
\subsection{Finesse}

The cavity array free spectral range (FSR) was acquired by slowly temperature scanning a Vescent $780 \nm$ DBR laser diode across $300 \GHz$. Cavity transmission was collected onto a \mbox{Thorlabs DET36A Si} \ac{PD} using a 10X magnification microscope objective lens. The signal was amplified using the $10^9$ low noise gain setting on a \mbox{Femto DLPCA-200} current amplifier. The cavity thermal mount was set to \mbox{$27\pm 0.1 \degree$C} and brought to thermal equilibrium.

The Rb \ac{SAS} signal near 780.24 nm \cite{steck_rb87} was recorded during the laser scan to provide a laser frequency reference for the diode laser temperature scan. The cavity spectra and Rb \ac{SAS} signal were measured simultaneously on an oscilloscope (\mbox{PicoScope 5242D}). Naming convention of the 12 cavities in the array followed a letter assignment (labeled A-L) as seen in the zoomed-in picture of the mounted cavity array in Fig \ref{fig:setupMount}b. Cavity K was chosen as the Rb \ac{SAS} was centered in the FSR for the mentioned set temperature. To confirm the detector was centered on the cavity of choice in the array, the \ac{PD} was replaced with an Imaging Source \mbox{DMK 33UX264} camera. The \ac{FOV} for this lens system spans approximately $800 \um$, which encompassed 3 adjacent cavities. An iris was placed near the imaging plane to isolate the central cavity in the \ac{FOV}. There was no measured increase in transmission signal associated with contribution from neighboring cavities with the iris fully open compared to the iris partially closed, when the cavity of interest was centered on the \ac{PD} active area (13 mm$^2$ in the image plane). 
\par

\begin{figure}[htbp]
\centering\includegraphics[width=\textwidth]{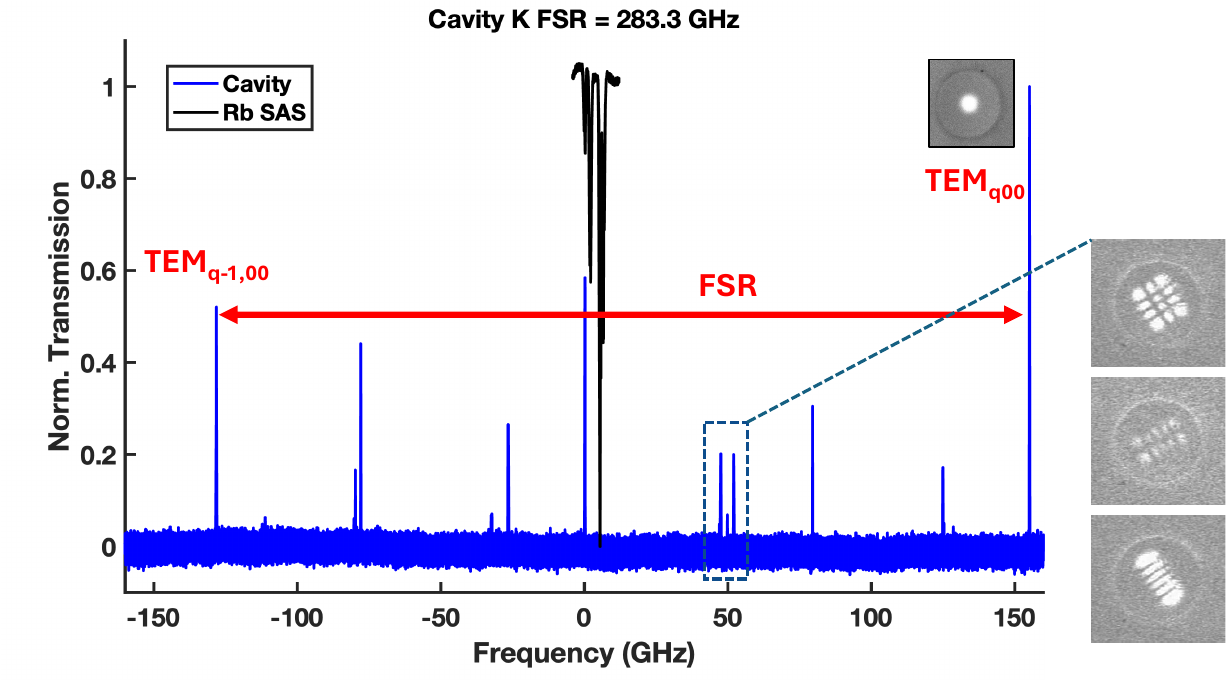}
\caption{Free spectral range (FSR) of Alluxa coated cavity measures 283.3 GHz. Scan is calibrated by Rb \ac{SAS} between $^{87}$Rb F=1--2’ and F=2--3’ transitions. Indicated on the plot are the sequential longitudinal TEM$_{00}$ modes of cavity K to where the FSR was measured. Imaged are a family of high order modes observed during the cavity resonance scan: TEM$_{q22}$, TEM$_{q14}$, TEM$_{q50}$ respectively. Due to the large span on the FSR, the Vescent DBR laser frequency was tuned by ramping the laser diode temperature with an external function generator at 25 mHz.}
\label{fig:finesse}
\end{figure}

The cavity transmission along with the Rb \ac{SAS} signal are shown in Fig. \ref{fig:finesse}. The resultant FSR = $283.3 \pm 0.05 \GHz$ was determined from the peak separation of the primary modes across one longitudinal mode index. Measurement uncertainty is dominated by the laser frequency stability in the thermal tuning configuration. Fig. \ref{fig:finesse} inset shows an example family of high order resonance modes taken on the camera. Their ready availability shows promise for higher-order atom interactions with modes where $m+n > 0$ \cite{du_precision_2013}.

\begin{figure}[htbp]
\centering\includegraphics[width=\textwidth]{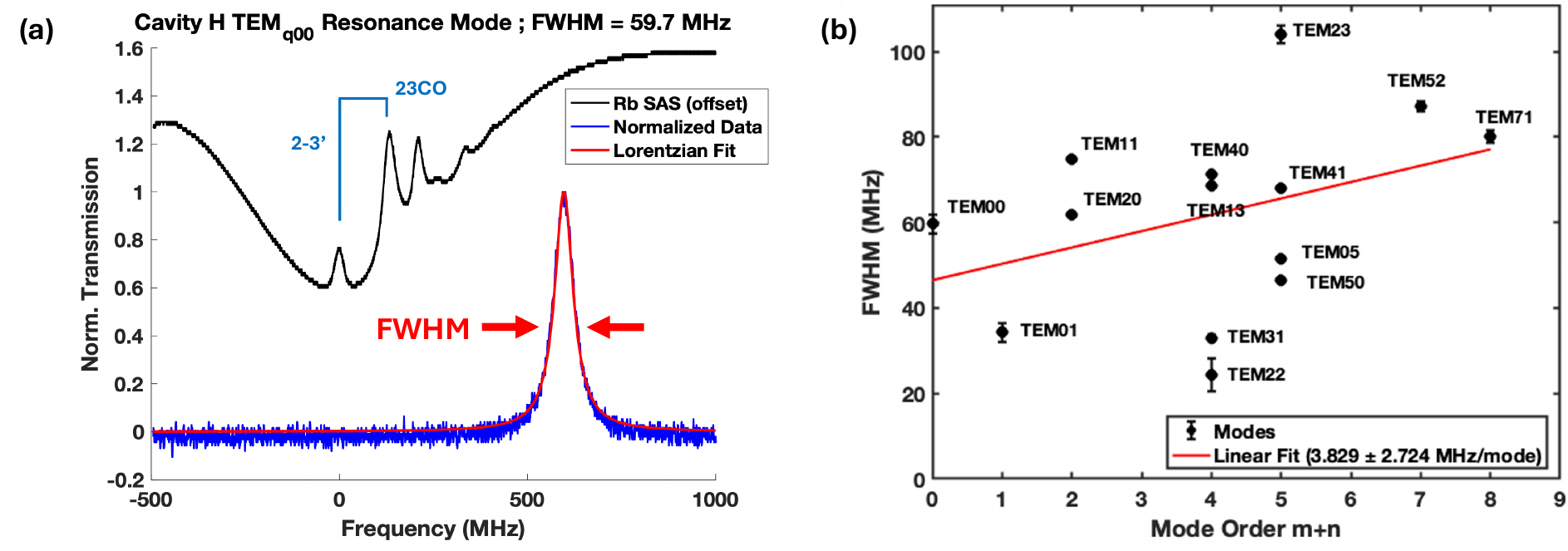}
\caption{(a) Lineshape measurement of cavity H TEM$_{q00}$ resonance mode is frequency calibrated using the 133.3 MHz spacing of the 2--3' and 23CO ($\lvert 5S_{1/2}, F=2 \rangle \rightarrow \lvert 5P_{3/2}, F=3 \rangle$ and $\lvert 5S_{1/2}, F=2 \rangle \rightarrow \lvert 5P_{3/2}, F=2\text{--}3 \rangle$ cross-over) transition lines found in the $^{87}$Rb \ac{SAS}. Laser current is scanned at 1 kHz. Lorentzian fit to resonance transmission data measures linewidth of 59.7 MHz. (b) Linewidth measurements of high order resonant modes within cavity L are grouped by degenerate m+n mode order.}
\label{fig:FWHM_HOM}
\end{figure}

%\par
The TEM$_{q00}$ resonance mode of cavity H (Fig.\ \ref{fig:FWHM_HOM}a) was analyzed by fitting the transmission across a 1 kHz laser current scan to a single Lorentzian with a measured linewidth of $59.7 \pm 2.5$ MHz. As shown in Fig.~5, cavities H and K are sufficiently close in resonance frequency that interchanging their linewidth measurements is reasonable when determining finesse. Treating the free spectral range (FSR) uncertainty as a fractional calibration error, an uncertainty of order $10^{-4}$ yields $< 1~\mathrm{MHz}$ for a $5~\mathrm{GHz}$ separation between cavities H and K. This is well below the measured resonance-frequency uncertainty of $\pm 114~\mathrm{MHz}$, indicating that use of the same FSR for both contributes negligibly to the overall error of the calculated finesse.

The cavity finesse, which is the ratio of the FSR to the resonance linewidth, is directly proportional to the cavity lifetime. 
Assuming mirror reflectivity is the dominant loss in the cavity, the finesse is related to reflectivity ($R$) according to:
\begin{equation}
    \mathcal{F} = \frac{\pi \sqrt{R}}{1-R} 
    \label{eq:finesseReflectivity}
\end{equation}
A lower bound for mirror reflectivity can be calculated from finesse value using Eq \ref{eq:finesseReflectivity} assuming negligible absorptive, scattering, and clipping losses.  Using this technique, the measured cavity finesse $\mathcal{F} = 4745 \pm 199$ corresponds to a reflectivity of approximately $99.93\%$, or equivalently $700 \pm 55$ ppm loss per mirror, in good agreement with the manufacturer's coating specification.

%==================================================================

\subsection{High Order Resonance Modes}

Application of high order resonance modes in \ac{CQED} has been demonstrated in \cite{du_precision_2013}, where tilted high order Hermite-Gaussian (HG) resonance modes provided unique insight to atomic motion through the asymmetric cavity mode. Higher order modes appearing in the cavity transmission scans correspond to different longitudinal index modes and are shifted due to the Gouy phase shift. The cavity stability parameters determine the order of transverse modes available within one FSR, which is noted by an integer increase of the longitudinal index \cite{siegman_1986lasers}:

\begin{equation}
    \frac{1}{\pi}(m+n+1) \cos^{-1}\sqrt{g_1g_2} \leq 1
    \label{eq:degenLim}
\end{equation}

\noindent where $g_n = 1-\frac{L}{\mathcal{R}}$ is the stability parameter while $m$ and $n$ are the transverse mode orders. The upper limit for transverse mode family degeneracy within an FSR may be estimated from Eq. (\ref{eq:degenLim}) to determine which $m + n$ transverse mode values are grouped in a longitudinal mode index of $q = N$ and $q = N+1$. Substituting cavity stability parameter values into Eq. \ref{eq:degenLim} results in $m + n \leq 3$. In Fig. \ref{fig:finesse}, values of $m+n\leq 3$ belong within index $q=N$, values of $4 \leq m+n\leq 6$ belong within index $q=N-1$.

\par

Higher resolution transmission spectra of all cavity resonances within a FSR of cavity L were collected using a narrow frequency scan centered about each resonance.  The data are plotted in Fig.\ \ref{fig:FWHM_HOM}.  To systematically measure all available modes, the central frequency of the scan was slowly adjusted while observing cavity transmission on the Imaging Source camera before swapping for the \ac{PD} and amplifier.  Frequency scanning was achieved through scanning the laser current; the cavity temperature was held at a constant value of $25 \pm 0.1 \degree C$.  A camera was used to identify the transverse order of each resonance.  The camera was then replaced with a \ac{PD} to measure the transmission spectra.  Lorentzian fits were applied to each transmission curve and calibrated against Rb spectra taken over the same $\sim 1$ GHz scan range. 

With the modes sorted by increasing $m+n$ transverse mode index, a linear fit to the data shows only a modest increase in FWHM by a factor of 1.6 over the range of cavities studied.  This increase is to be expected for higher order modes that experience greater diffractive loss \cite{siegman_1986lasers}.  The deviation from the trendline is even more significant, however.  For instance, the TEM 22 and 23 modes differed by a factor of 5 in FWHM though the mode index increased by only 1.  We speculate this could be due to the fine details of the mirror etching profile, for which more exploration is needed.

%==================================================================
\begin{figure}[htbp]
\centering\includegraphics[width=\textwidth]{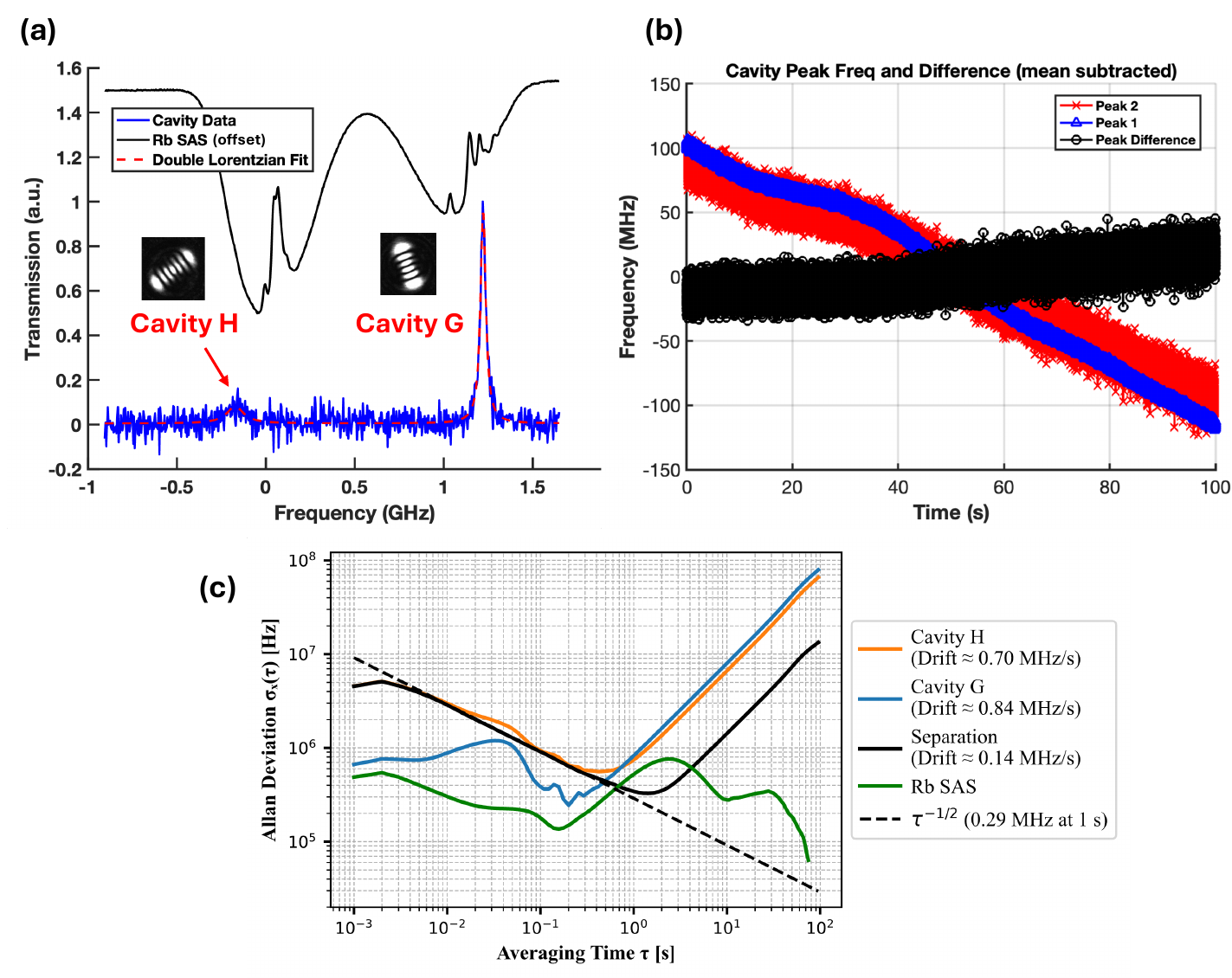}
\caption{(a) Co-resonant modes TEM$_{q06}$ and TEM$_{q50}$ in cavities H and G respectively frequency calibrated using Rb absorption spectra. (b) Mean subtracted fitted resonance peak frequency for each cavity and the mean subtracted frequency difference between each peak. Data frequency calibration factor is $5.86 \pm 0.03 ~MHz/\mu s$. Cavity was mounted using a copper capping plate and damped steel screws on a temperature-controlled mount. The temperature controller was off for this dataset, leaving the cavity at $23.1 \pm 0.1 \degree \mathrm{C}$. (c) Overlapping Allan deviation of cavity peak times, peak difference, and Rb \ac{SAS}.}
\label{fig:cavGH}
\end{figure}

\subsection{Stability of Co-Resonant Cavities}

We explored the relative passive stability of the cavities in the configuration with a flat, stable spacer element inserted between the two cavity mirrors.  To determine the relative stability of adjacent cavities, the resonances of neighboring cavities must be monitored simultaneously. Towards this end, a 2 mm diameter beam was used to couple light into multiple adjacent cavities, while a camera imaging system monitored the transmission from neighboring cavities. The challenge was to simultaneously observe two resonances whose absolute frequency difference was within the laser frequency sweep range of 2.5 GHz in {\em different} cavities.  Fig.\ \ref{fig:cavGH} shows data for a cavity pair with nearly degenerate TEM$_{06}$ and TEM$_{50}$ modes.

Co-resonant cavity spectra in cavities G and H were taken with the sample  in an ordinary room temperature environment with no attempt made to stabilize the temperature ($T=23.1 \pm 0.1 \degree$C). %The data from cavity H showed a larger variance due to the lower light level observed at the peak seen in Fig.\ \ref{fig:cavGH}a.  
Cavity H showed a larger variance compared to cavity G in the measured resonance peak position due to the lower transmitted light level and reduced SNR seen in Fig.\ \ref{fig:cavGH}a.  Nonetheless, in Fig.\ \ref{fig:cavGH}b it is easy to pick out a common drift between the cavities associated with slowly increasing lab temperature. Overlapping Allan deviations of this data (Fig. \ref{fig:cavGH}c) show a strong white noise component in each cavity resonance position, evidenced by the $t^{-1/2}$ scaling at short times up to about 0.5 seconds.  For longer times the cavity positions drifted, likely due to sample temperature drifts in the lab environment.  The data in Fig.\ \ref{fig:cavGH}c have been fit to a $\sigma (\tau) = A/\sqrt{\tau} + B\tau$ model from which we could extract the signal drift rates ($B$). We estimate that for a $500~\mu$m silicon spacer with coefficient of thermal expansion $\alpha = 2.6$ ppm/$\degree$C the observed drift rates correspond to temperature changes $<0.001\degree$C/s, using the formula $\frac{\delta \nu}{\nu} = -\alpha ~\delta T$.  Strong suppression of common mode fluctuations was observed, however, in the cavity difference data set, whose Allan deviation showed a factor of about 5 reduction in drift rate to 0.14 MHz/s.  Small temperature gradients or minor stress differences could be responsible for this residual drift rate.  Future work will utilize tight temperature control ($0.001$ K is straightforward) to explore the limits of passive stability that are achievable.

%==================================================================

\begin{figure}[htbp]
\centering\includegraphics[width=\textwidth]{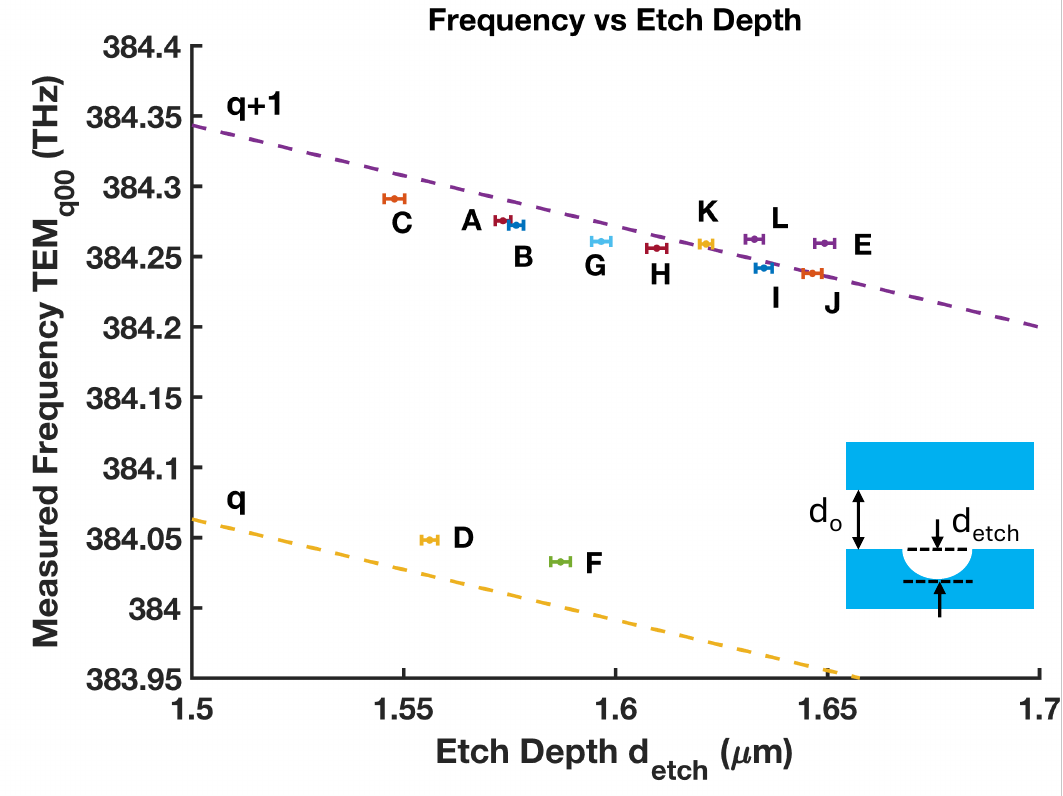}
\caption{Measured absolute frequency of $TEM_{q00}$ mode for each cavity in the array against etch depth. Resonance mode frequency data was collected by scanning the input laser frequency, via thermal tuning the laser diode. Profilometry measurements of each cavity etch depth with horizontal error bars showing RMS error for each unique cavity were performed at AFRL. Dashed lines indicate theoretical TEM$_{q00}$ resonance mode given best fit longitudinal index value (q) and measured silicon spacer thickness ($d_o = 533.3\um$) with a fixed radius of curvature $\mathcal{R}$ of --1 mm.}
    \label{fig:alluxaEtch}
\end{figure}

\subsection{Array Etch Depths}

The absolute resonance frequency of each cavity is a highly variable quantity, depending on not just the cavity etch depth but also tiny variations in the spacer thickness and alignment.  For future batch fabricated systems, it is crucial to understand this variability and how it can be reduced.  While no attempt was made to control the machined depth of the mirrors in the present work, we could nonetheless explore the relationship between the measured cavity depths and resonance frequencies, thus isolating the effects of spacer alignment.  
 
When analyzing our cavity primary resonance mode (TEM$_{qmn}$), we may substitute $m=n=0$ and express cavity length as $L=d_0 + d_{etch}$, where $d_0$ is the spacer thickness and $d_{etch}$ is the individual cavity etch depth as shown in the inset in Fig. \ref{fig:alluxaEtch}. The TEM$_{q00}$ resonance of a cavity may be calculated as \cite{siegman_1986lasers}:

\begin{equation}
    \nu_{q00} = \frac{c}{2(d_0+d_{etch})} \left( q + \frac{1}{\pi} \cos^{-1}\sqrt{1-\frac{d_0+d_{etch}}{\mathcal{R}}} \right)
    \label{eq:resFreq}
\end{equation}
\noindent The TEM$_{q00}$ mode of each cavity in the array was captured on video at 30 frames per second with the camera during a 5 mHz ramp scan of the input laser frequency. Simultaneous measure of the absolute input laser frequency was monitored using a Bristol wavelength meter. The optical cavity was thermally stabilized at $25 \pm 0.1\degree$C. All cavities in the array scanned through one FSR encompassing two TEM$_{q00}$ modes. The video scan ramp range was found with $\pm 33$ ms uncertainty, corresponding to one video frame. Calibration of the frequency step per frame for the data set results in measured frequency uncertainty of $\pm 114$ MHz. The measured TEM$_{q00}$ resonant frequencies plotted against each etch depth are shown in Fig. \ref{fig:alluxaEtch}, with straight lines corresponding to Eq.\ \ref{eq:resFreq}.  There is a strong correlation between the two, which suggests that refinements to the etch process to control the depth is a viable path forward to achieving degeneracy of longitudinal mode resonances between neighboring cavities.   

%==================================================================

\section{Discussion and Conclusion}

Our results show the potential for designing multiple co-resonant cavities using the approach detailed in this work.  The reduction of cavity mode frequency spread from $\sim 100$ GHz down to $\sim 1$ GHz through finer control of relative etch depths in the fused silica substrate may be possible in order to realize this. With this control, the need for micro-piezo actuators for individual fine control may be greatly reduced, restricted only to adjusting for small noise contributions such as thermal noise or package stress.  Our studies of passive stability, moreover, show promise for the use of cavity arrays for applications where the absolute frequency of transmission between neighboring cavities can be controlled.  It is also possible to design an array of etch depths with an intentional frequency offset within the cavity array, which would be useful in targeting other high order resonance modes that overlay a Rb resonant input frequency.

\par

One potential application is in a chip-scale atom-optical interaction platform that utilizes an array of micro-\acp{FPC} each aligned to a collimated Rb atom beam. A co-packaged Rb atom beam source would provide a continuous stream of atoms to each micro cavity in the array. Both the collimated atom beam and compact cavity platform, designed for compatibility with MEMS fabrication techniques, are advancements towards realizing scalable and even perhaps one day, portable cavity quantum electrodynamics (CQED) platforms.

%==================================================================

\begin{backmatter}
\if0
\bmsection{Funding}
\textcolor{orange}{Content in the funding section will be generated entirely from details submitted to Prism. Authors may add placeholder text in this section to assess length, but any text added to this section will be replaced during production and will display official funder names along with any grant numbers provided. If additional details about a funder are required, they may be added to the Acknowledgment, even if this duplicates some information in the funding section. For preprint submissions, please include funder names and grant numbers in the manuscript.}
\fi

\bmsection{Acknowledgment} Funding provided by the United States Space Force and the Air Force Research Laboratory under Universities Space Research Association Cooperative Agreement Number FA9453-21-2-0064. 

\bmsection{Disclaimer}                                                           
The views expressed are those of the authors and do not necessarily
reflect the official policy or position of the Department of the Air
Force, the Department of the Defense, or the U.S. Government.

\bmsection{Data Availability Statement} Data underlying the results presented in this paper are not publicly available at this time but may be obtained from the authors upon reasonable request.

\bmsection{Supplemental document}
See Supplement for supporting content.

\end{backmatter}

%%%%%%%%%%%%%%%%%%%%%%% References %%%%%%%%%%%%%%%%%%%%%%%%%

%Add references with BibTeX or manually 
%\cite{Zhang:14,OPTICA,FORSTER2007,Dean2006,testthesis,Yelin:03,Masajada:13,codeexample}.

\bibliography{sample}

\end{document}

\maketitle
\thispagestyle{fancy}

% ==============================================
\section{Silicon spacer fabrication}
\begin{figure}[htbp]
\centering
\includegraphics[width=\textwidth]{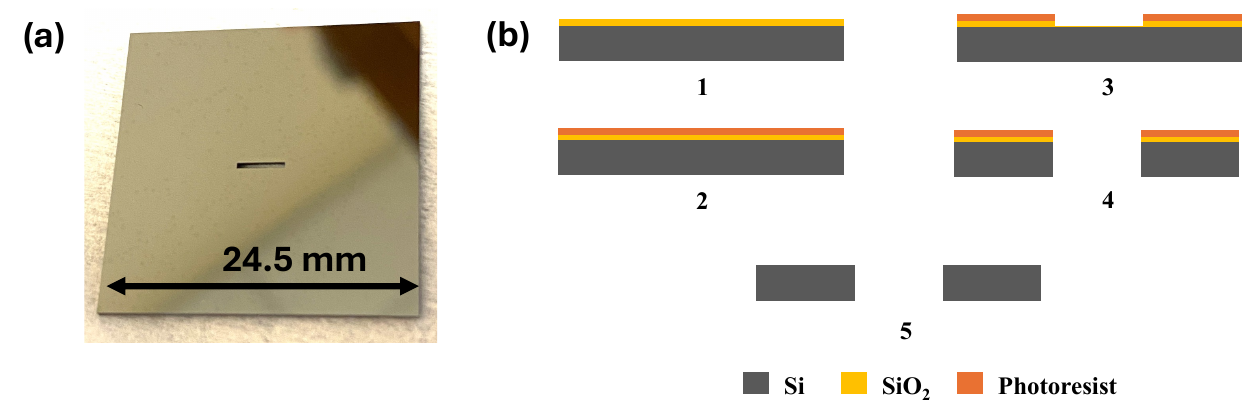}
\caption{(a) Fabricated silicon spacer (1" x 1") for micro-cavity array. Through hole etched into substrate (0.5 mm x 4 mm) aligns with position of etched cavity array on fused silica substrate to allow for cavity transmission measurements. (b) Process flow for silicon spacer: (1) deposition of SiO$_2$ onto silicon substrate, (2) spin coating photoresist, (3) plasma etching lithographically defined pattern into oxide layer, (4) DRIE silicon, and (5) removal of mask layers with piranha and BOE.}
\label{fig:spacerFab}
\end{figure}

To ensure maximum mechanical stability, the silicon spacer was designed to provide maximum surface contact with the cavity array glass slides.  To provide optical access to the cavities, a small slit was cut into the silicon spacer (see Fig.~\ref{fig:spacerFab}).  To ensure flatness of the fabricated surface, a chemical etching process was used.

A $2~\mu$m layer of silicon dioxide was deposited on a 3-inch diameter, single-side polished $400~\mu$m thick silicon wafer using plasma-enhanced chemical vapor deposition (PECVD). A $5~\mu$m-thick layer of positive photoresist (SC 1827) was then spin-coated at 4000 rpm for $40 \s$ and soft-baked at 120 $\degree$C for 3 minutes. The wafer was patterned using a $405 \nm$ laser exposure at $275 \mJ/\cm^2$, followed by development in MF-319 for 3 minutes. The exposed oxide was removed by plasma etching, after which deep reactive-ion etching (DRIE) was performed to define the desired features. Finally, the wafer was cleaned sequentially in piranha solution and buffered oxide etch (BOE).

%==================================================================

\section{Etch Radii of Curvature and birefringence}

Polarization degeneracy in the cavity resonances may break due to birefringence of dielectric mirror coating, anisotropic cavity etching, or through stresses in the mirror substrates.  The cavity did not show dependence on input light polarization. Cavity transmission scans were taken over the cavity L $TEM_{q00}$ mode and cavity L $TEM_{q01}$ mode with vertical, horizontal, and 45 degree input light polarizations. The input polarization was also continuously tuned between horizontal and vertical polarization while observing the resonance mode to observe if any power exchange between peaks or change in lineshape occurred. The varied input polarization measurement was repeated after rotating the flat mirror by 90 degrees. No changes were seen for either mirror orientation, indicating the cavity is insensitive to input polarization.  
\par
\begin{figure}[htbp]
\centering
\includegraphics[width=\textwidth]{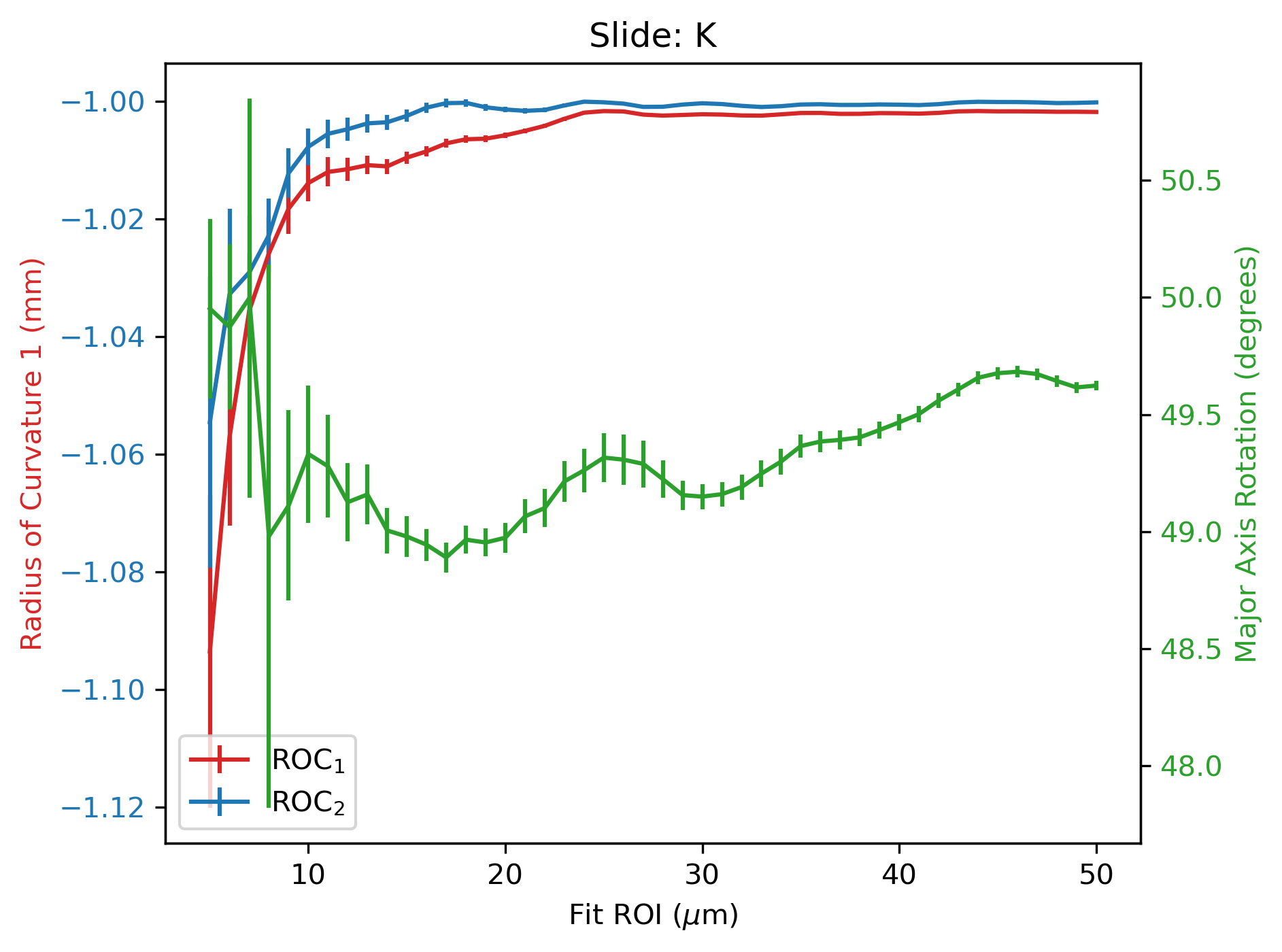}
\caption{Result of a least-squares fit of surface height data for
cavity K.  Height data was fit to a parabolic surface with
independently parameterized radii of curvature along orthogonal
major and minor axes; radii of curvature and major axis rotation
are free variables of the fit.  Since resonant beams have
transversely varying intensity at the mirror surface, the fit was
repeated for height data restricted to different region of
interest (ROI) radii about the mirror center.  Similar radii of curvature are seen along the major and minor axes for ROI radii above 10~$\mu$m.  Error bars show least-squares fit error.}
\label{fig:etchROC}
\end{figure}

Fabricated microcavities using CO$_2$ laser etching results in a major and minor axis in the mirrors, as seen in Ref.~\cite{uphoff_2015}. The eccentricity of the resulting elliptical features lifts the degeneracy in the polarization eigenmodes. The impact of mirror eccentricity becomes non-negligible when the radius of curvature is on the order of the resonant wavelength \cite{uphoff_2015}. The mirror radius of curvature is --1 mm in our cavity array, as seen in Fig. \ref{fig:etchROC} for cavity K. Since the etch radius of curvature is much larger than the resonance wavelength, $\lambda \sim 780$ nm, we do not expect to observe frequency splitting in our mode.

% ==============================================
\section{Array Etch Depths}

A cavity optical mode is resonant when the field is the same phase after a round trip ($2L$) inside a cavity. This means that the phase evolution over the round trip is a multiple of $2\pi$. The phase may be expressed as $\phi(z) = -kz + \xi(z)$, where $\xi(x)$ expresses the axial phase difference due to the field propagation through a focus (known as the Guoy phase shift). The total phase shift for the field propagating from one end of the cavity to the other ($L$) can be equated to integer multiples ($q$) of $\pi$ such that $\Delta \phi = \phi(z_2 - z_1) = -q \pi$.

\par 

For Hermite-Gaussian modes (TEM$_{qmn}$), the total phase shift across the cavity, including the Guoy phase shift terms, for an $mn^{th}$ order mode is expressed as:

\begin{equation}
    -q \pi = k(z_2 - z_1) - (m+n+1)\{ \xi(z_2)-\xi(z_1) \}
\end{equation}

\noindent Here we can know $k(z_2 - z_1) = kL$, where $L$ is the length of the cavity in one direction of transit, and we can also substitute $k=2\pi\nu/c$ into the expression to solve for frequency ($\nu$). The Guoy phase shift terms are related to the Gaussian beam parameters by: $\xi(z_i)=\tan^{-1}\left(z_i/z_R\right)$, where $z_R$ is the Rayleigh range of the beam. The Guoy phase term can be expressed using the cavity stability parameter $g_i = 1-(L/\mathcal{R}_i)$ for each mirror, where the radius of curvature of each mirror can also be expressed using the Gaussian beam parameters $\mathcal{R}_i = z_i + (z_R^2/z_i)$. The frequency of TEM$_{qmn}$ can be written using the cavity stability parameters as follows: 

\begin{equation}
    \nu_{qmn} = \frac{c}{2L} \left( q + \frac{1}{\pi}(m+n+1) \cos^{-1}\sqrt{g_1g_2} \right)
    \label{ResFreq}
\end{equation}

\noindent Our microcavity array is plano-concave (PC), meaning the stability parameter for the flat and curved mirrors are respectively: 

\begin{equation}
    g_1 = g_{flat} = 1
\end{equation}
\begin{equation}
    g_2 = g_{etch} = 1-\frac{d_0 + d_{etch}}{\mathcal{R}}
\end{equation}

\noindent where the cavity length is expressed as the sum of the silicon spacer ($d_0$) and the etch depth ($d_{etch}$) of mirror 2. The silicon spacer thickness was measured with micrometers as $541 \pm 2.54 ~ \mu$m thick. 

% ==============================================
\section{Adjusted control and acquisition parameters for cavity stability}

\par
Cavity spectra data were collected by performing
a linear laser frequency sweep while collecting
photodiode data on an oscilloscope.  When using
this technique, it is important to link the time
domain of the oscilloscope to the frequency of
the laser. For a linear frequency scan, time and
frequency are related by a constant
multiplicative calibration factor, k. A saturated absorption spectroscopy signal (SAS) of Rb
provided an absolute frequency reference. 
Cavity transmission data and SAS signal were
simultaneously recorded on the same oscilloscope.
SAS data near the $|5\mathrm{S}_{1/2}, \mathrm{F}=2\rangle \rightarrow |5\mathrm{P}_{3/2}, \mathrm{F}=3\rangle$
and
$|5\mathrm{S}_{1/2}, \mathrm{F}=2\rangle \rightarrow |5\mathrm{P}_{3/2}, \mathrm{F}=2-3\rangle$ cross-over peaks were least-squares fit to a
quadratic function.  The calibration factor, $k$, was calculated by
dividing the known frequency separating the resonances in Rb by the
measured time separating the resonances on the oscilloscope. 
Uncertainty in the timing of the resonances is estimated from the
least squares fit.  Assuming uncorrelated error, the uncertainty in
the time separating the resonances may be calculated as $\sigma_{\Delta
t}^2 = \sigma_1^2 + \sigma_2^2$, where $\sigma_1$ and $\sigma_2$ are the timing uncertainties of the two resonances.

\par
For the co-resonant cavities dataset, the peaks of each cavity resonance mode was tracked over a series of ramps to determine their individual positions and relative spacing in time along with their associated errors.
Due to the comparatively weak transmission signal seen in cavity H, the following scheme was used to reliably fit to both cavity resonances.
An initial double Lorentzian fit was applied to the data to determine the nominal spacing between the two peaks, giving an estimate time separation of $\Delta t_{avg}$. For each subsequent ramp, the taller peak was fit with a single Lorentzian to find its timing $t1$ and fit error ($\sigma_{t1}$). With the first peak timing found and the nominal spacing between the peaks extracted, a double Lorentzian was applied to each respective scan. A small range bounding the expected second peak position ensured a robust process for fitting low SNR peak signals, yielding the second fit peak timing $t2$ and fit error $\sigma_{t2}$.  Timing separation between the two cavity resonances is converted into the frequency domain using the calibration constant, $k$.  Uncertainties in both peak frequencies, along with the uncertainty in the calibration factor, are combined through propagation of errors to determine the uncertainty in the frequency separating the two cavity resonances; all errors are assumed to be uncorrelated.

\begin{figure}[htbp]
\centering
\includegraphics[width=\textwidth]{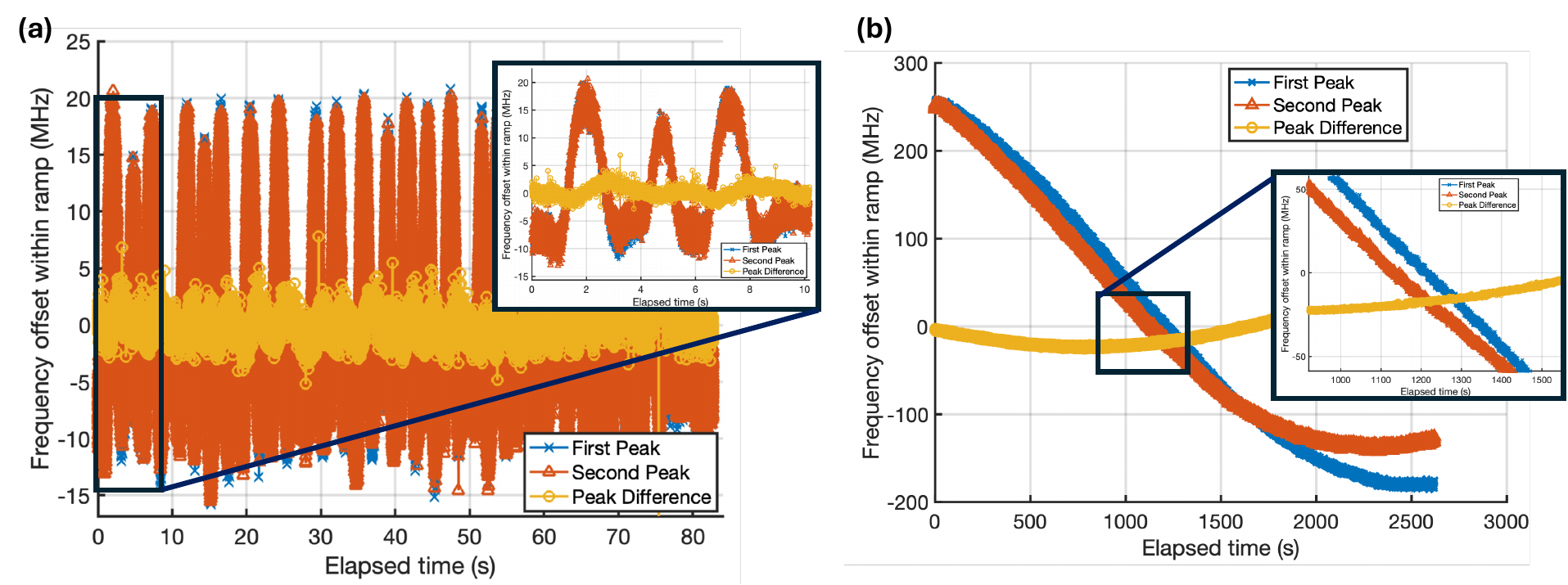}
\caption{(a) Mean subtracted fitted transition peaks in Rb saturated absorption spectra (SAS) for $^{87}$Rb $|5\mathrm{S}_{1/2}, \mathrm{F}=2\rangle \rightarrow |5\mathrm{P}_{3/2}, \mathrm{F}=3\rangle$
and
$|5\mathrm{S}_{1/2}, \mathrm{F}=2\rangle \rightarrow |5\mathrm{P}_{3/2}, \mathrm{F}=2-3\rangle$ cross-over. Scans of spectra occur every 1 ms. Dataset frequency calibration $5.64 \pm 0.27 ~MHz/\mu s$. Inset zooms in on oscillations occurring in spectra and incomplete common mode rejection in peak difference data. (b) Mean subtracted fitted transition peaks between $^{87}$Rb $|5\mathrm{S}_{1/2}, \mathrm{F}=2\rangle \rightarrow |5\mathrm{P}_{3/2}, \mathrm{F}=3\rangle$
and
$|5\mathrm{S}_{1/2}, \mathrm{F}=2\rangle \rightarrow |5\mathrm{P}_{3/2}, \mathrm{F}=2-3\rangle$ cross-over, for 1 kHz scan taken every 100 ms. Dataset frequency calibration $5.39 \pm 0.31 ~MHz/\mu s$. Inset zooms into data shows drift of frequency but no oscillations.}
\label{fig:RbSAS}
\end{figure}

\par
Comparison of the cavity and peak difference datasets against the Rb saturated absorption spectra (SAS) in the Allan deviation indicate that the cavity noise variation is not limited by the input laser. A bump in the Rb SAS data is observed at $\tau \sim 2$ s. Data of tracked $^{87}$Rb $|5\mathrm{S}_{1/2}, \mathrm{F}=2\rangle \rightarrow |5\mathrm{P}_{3/2}, \mathrm{F}=3\rangle$
and
$|5\mathrm{S}_{1/2}, \mathrm{F}=2\rangle \rightarrow |5\mathrm{P}_{3/2}, \mathrm{F}=2-3\rangle$ cross-over transition peaks (Fig. \ref{fig:RbSAS}a) show an $\sim 0.3$ Hz oscillation occurring in the spectra, which correlates to the seen bump in the Allan deviation at half the sine period. For a sinusoidal signal with period $T$, averaging over interval length $\tau = T$ will cover a full oscillation cycle so that the mean value over consecutive averaging windows have small variations. However, for $\tau = T/2$ each average covers half the oscillation cycle so that consecutive averages show maximum differences. The oscillations seen in the 1 kHz acquisition rate data set are not seen in the 10 Hz acquisition rate data (Fig. \ref{fig:RbSAS}b taken at 1 kHz scan rate in 100 ms intervals). The underlying cause for the difference in Rb SAS stability between the 1 ms and 100 ms acquisition intervals lies with the response of the laser diode temperature control PID. At a continuous 1 kHz scan rate of the laser diode current, the diode temperature is not effectively stabilizing which results in a variation of the central frequency. In contrast, the laser diode current scan settings for the 100 ms interval data utilizes burst mode in the Agilent function generator. This allows a 1 kHz scan rate to be gated so that it occurs every 100 ms. Here, wait times are not scanning the diode current, which allows better stabilization of the laser diode temperature.

% ==============================================

% Bibliography
\bibliography{suppl}